\documentclass[]{elsarticle}

\usepackage{cite}
\usepackage{graphicx}
\usepackage{xcolor}
\usepackage{url}
\usepackage{amsmath, amsthm, amssymb}
\usepackage{booktabs}
\usepackage{tabularx}
\usepackage{todonotes}
\usepackage[normalem]{ulem}

\makeatletter
\def\ps@pprintTitle{%
 \let\@oddhead\@empty
 \let\@evenhead\@empty
 \def\@oddfoot{}%
 \let\@evenfoot\@oddfoot}
\makeatother

\newcommand{\correction}[1]{\textcolor{black}{#1}}
\newcommand{\correctionbis}[1]{\textcolor{black}{#1}}
\newcommand{\delay}{\ensuremath{\overline{\delta t}}}

\newcolumntype{L}[1]{>{\raggedright\let\newline\\\arraybackslash\hspace{0pt}}m{#1}}
\newcolumntype{C}[1]{>{\centering\let\newline\\\arraybackslash\hspace{0pt}}m{#1}}
\newcolumntype{R}[1]{>{\raggedleft\let\newline\\\arraybackslash\hspace{0pt}}m{#1}}

\graphicspath{{new_figures/}}

\begin{document}

\title{The economic value of additional airport departure capacity}

\date{\today}

\author[westminster]{G\'{e}rald Gurtner\corref{correspondingauthor}}
\author[westminster]{Anne Graham}
\author[westminster]{Andrew Cook}
\author[innaxis]{Samuel Crist\'{o}bal}

\address[westminster]{School of Architecture and Cities, University of Westminster, 35 Marylebone Road, London NW1 5LS, United Kingdom}
\address[innaxis]{The Innaxis Foundation and Research Institute, Calle de Jos\'{e} Ortega y Gasset, 20, 28006 Madrid, Spain}

\cortext[correspondingauthor]{Corresponding author}

\begin{abstract}

This article presents a model for the economic value of extra capacity at an airport. The model is based on a series of functional relationships linking the benefits \correctionbis{of} extra capacity and the associated costs. It takes into account the cost of delay for airlines and its indirect consequences on the airport, through the loss or gain of aeronautical and non-aeronautical revenues. The model is highly data-driven and to this end a number of data sources have been used. In particular, special care has been used to take into account the full distribution of delay at the airports rather than its average only. The results with the simple version of the model show the existence of a unique maximum for the \correctionbis{operating profit} of the airport in terms of capacity. The position of this maximum is clearly dependent on the airport and also has an interesting behaviour with the average number of passenger per aircraft at the airport and the predictability of the flight departure times. \correctionbis{In addition, we} also show that there exists an important trade-off between an increased predictability and the punctuality at the airport. Finally, it is shown that a more complex behavioural model for passengers can introduce several local maxima in the airport \correctionbis{profit} and thus drive the airport towards suboptimal decisions.\footnote{\url{https://doi.org/10.1016/j.jairtraman.2018.01.001}}

\end{abstract}

\maketitle

\section{Introduction}

A number of major airports in Europe are already under stress due to high volumes of traffic during peak times \citep{gelhausen}. Since traffic in Europe is expected to grow by 50\% in the next 20 years \citep{Eurocontrol2013}, it is expected that many other airports will be severely congested in the medium term, and that airports that are currently congested at peak times will have problems all day long. As a consequence, \correctionbis{the major European public-private research partnership SESAR (Single European Sky ATM Research)} has dedicated an Operational Focus Area (OFA05.01.01) to the development of the Airport Operations Center (APOC) to consider mitigation measures to avoid large delays at these airports and the associated costs.

Delays are a direct consequence of levels of congestion at airports. These impact directly on the airlines. For these, delays usually mean sub-optimal levels of operation, as well as decreased satisfaction of their customers, leading to potential decreases of market share. The value of this shortfall can be evaluated for different types of airline, aircraft, and \correctionbis{delay duration}, etc. \citep{cost_delay}.

However, it is clear that expanding the capacity of an airport is costly. Depending on the nature of the bottleneck and the severity of the congestion, the airport might need to physically expand its infrastructure. This could mean, for example, increasing the number of runways, the number of terminals, or the number of gates. In all cases, the \correctionbis{total \textit{operating}} costs for the airport will be higher after the expansion. As a result, there will be an optimal capacity for the airport which balances the level of congestion with the costs associated with the extra capacity.

This is the concept which is explored in this paper, using a simple model to capture this effect. More specifically, the model aims \correctionbis{to provide} some quantitative measures of the cost of capacity and the corresponding cost of delay in a very data-driven way. To this end, different types of data have been collected that guide the modelling process and allow for detailed calibration. 

The structure of the paper is as follows. Section 2 presents the literature review, focusing on the main mechanisms that should be included in the model. The types and sources of data used are also discussed. Section 3 presents the model in detail, including the calibration process. Section 4 provides some results obtained with the model. Finally, conclusions are drawn in Section 5. 

\section{State of the art}

\subsection{Literature review}

Many studies have been undertaken concerning various aspects of airport economics over the past few years and in this section a concise overview of the most relevant research is provided. In particular, consideration is given to the main mechanisms that link capacity to cost and delay, and the associated strategies adopted by airports over the years.

\correctionbis{Since a significant part of an airport's operating costs \correctionbis{is} fixed, excess capacity will produce high overall unit costs, as the fixed costs will be spread over lower than optimal traffic levels. Whilst attempts may thus be made to use the current facilities as much as possible, to take advantage of economies of density or capacity utilisation \citep{mccarthy}, being close to capacity is likely to produce more delays. So both capacity utilisation and delays can have an impact on airport cost efficiency \citep{pathomsiri}, with \citep{adler2} empirically finding that the positive impact of utilisation is greater than the negative impact of delays.} 

Delays have impacts for both passengers and airlines \citep{cost_delay}. As passenger satisfaction may be linked to commercial spend \correctionbis{-- the money spent by passengers --} at the airport \citep{pax_satisfaction}, delays can have a direct negative impact on an airport's performance, although this relationship is yet to be confirmed  \citep{merkert} due to very limited research. \correctionbis{This in turn is} due to the lack of appropriate and publicly available passenger satisfaction data. On the other hand, higher delays at the airport may have the opposite effect, since passengers have more time to use the commercial facilities \citep{alfonso}, even though the only known empirical study in this area found no significant relationship between commercial revenues and delayed flights \citep{fuerst}.

Adapting airport capacity to the expected level of traffic is a complex task and many possibilities are discussed in the literature. First, so-called `soft' management approaches have been examined. These include minor modifications to management processes at the airport, without having an impact on the infrastructure itself. They are quick to implement and relatively low cost, but clearly limited in scope. They can relate to strategic planning or tactical adjustments \citep{barnhart}. They can also include more local solutions, such as improvement planning \citep{daniel,jorge}, changes to air traffic control (ATC) rules, price changes, and incentive schemes for airlines to use larger aircraft -- given that the infrastructure for this is already in place -- even if this may lead to additional congestion in the terminals \citep{gelhausen,berster}. In the broader sense, they include developing intermodality with high-speed trains, diverting traffic or using multi-airport systems \citep{martin}, even though these typically require at least some infrastructure change.

The feasibility and effectiveness of using pricing to manage congestion has been frequently discussed in the literature, with the theoretical arguments summarised by \citep{zhang}. However, such practices have rarely been applied and tested. One of the key issues is the extent to which airlines already self-internalise congestion, on which point views vary \citep{brueckner}. \correctionbis{Moreover, \citep{adler2}  empirically found that delays had no impact on aeronautical revenues but that this was significantly higher at congested airports}. Other research has shown that it is important to take into account different passenger types when assessing the efficiency of any potential new pricing scheme. Unsurprisingly, passengers having a higher value of time -- typically corresponding to business-purpose passengers -- will benefit from increased charges during peak times to protect them from the congestion caused by passengers with lower values of time \citep{czerny,yuen}. Such pricing solutions are also difficult to implement because many airports are subject to economic regulation, most commonly in the form of a price-cap \citep{adler}. Another alternative, but related, demand-management technique frequently studied in the literature is a type of reform of the current slot allocation process, for example by using slot auctions and secondary trading systems. This would have a major impact on airlines and passengers, but most likely a lesser impact on airport revenues \citep{madas,verhoef}.

The second possibility to cope with excess demand is to change the infrastructure itself, usually by extending the current number of terminals, runways, gates, etc.: so-called `hard' management approaches. These measures are usually slow to implement and very costly, but can bring great increases in capacity in some way or another. \correctionbis{There will be a significant lag between the potential expansion decisions and the full released capacity, during which demand and the environment may change. This introduces a complex dynamic behaviour of development and investment, which in part creates a demand for more flexible solutions \citep{leucci,kwakkel}. It also poses the problem of the risk aversion of the airport operators, and, more generally, the problem of how expectations are formed with regard to the likely investment return. Some research points out that the various uncertainties in the airport system, including the uncertainty of future demand \citep{xiao} and the unpredictability of degradation \citep{desart}, increase the difficulties of airport capacity decision-making processes. Moreover, as airports are not isolated entities, airline network (delay propagation) effects can add further complexity to the validity of a capacity extension \citep{cost_delay}. The decision-making process of the airport under various uncertainties is a complex subject, as noted in \citep{sun, kincaid}.}

The literature also points out the need for more subtle definitions of capacity, in particular ensuring that there is differentiation between arrival versus departure capacity, and runway versus terminal capacity. It has been shown that there \correctionbis{is} some trade-off between the former \citep{gilbo}, and that there exist some non-trivial relationships between the latter \citep{wan}. Currently, \correctionbis{runways typically} represent the bottleneck for the traffic flow, rather than terminals \correctionbis{\citep{gelhausen,berster,wilken,butler}}. There is also the trade-off between operational and commercial capacities, the extent of complementarity between these two, and the associated cost allocation approaches \citep{zhang_zhang,alfonso}. This is linked to the flexibility allowed within each individual airport economic regulatory system and subsequent incentives which may arise \citep{expanding}.

A common research theme concerns cost-benefit analyses examining the implications of a `hard' modification. In particular, it is important to emphasise that changing infrastructure may not merely affect the volume of traffic or passengers, but also the \correctionbis{nature of the} traffic and operations at the airport. Indeed, larger airports are usually more diversified in being able to provide a greater range of commercial facilities. As a consequence, commercial spend can increase disproportionately with the size of the airport. Also, leisure passengers have been shown to spend more than business passengers \citep{fuerst,castillo}, and low-cost carrier (LCC) passengers less \citep{lei}. Traffic mix changes will also bring different associated costs related to the service expectations of the airlines, related, for example, to ensuring a fast transfer time at hub airports, or swift turnarounds for LCCs.  As regards airport size, much mixed evidence exists, but generally it shows that airports experience \correctionbis{cost} economies of scale, albeit with different findings related to if, and when, these are exhausted, and whether diseconomies then occur. 
\correctionbis{For UK airports some research has estimated that long-run average costs decreased up to 5 million passengers, were constant for 5-14 million passengers, and then started to increase \citep{bottasso}, whereas another UK study \citep{main} found a steep decrease in average costs until around 4 million passengers and then very moderate, but persistent decreases in costs until at least 64 million passengers. Meanwhile, for Spain it has been concluded that cost economies are not exhausted at any level of traffic for the airports considered \citep{martin2011}, with similar results confirmed for a worldwide sample \citep{voltes}. These studies considered both operating and capital (i.e. long-run) costs.}

\correctionbis{A key related issue is how aeronautical charges may change as the result of the costs of new infrastructure.} However, it has been shown that aeronautical revenues are very much influenced by market-oriented factors, such as price sensitivity or competition \citep{bel,bilotkach}, as well as pure cost drivers. The impact of changes in charges may also be limited, since they \correctionbis{tend to} represent a small portion of the airline costs. \correctionbis{This} also depends on the extent to which airlines will absorb such changes or pass them on fully to passengers \citep{starkie}, which is difficult to evaluate without further empirical evidence.
\\

This literature review has provided a high-level overview of the airport system, in particular with regard to relevant variables and the relative importance of the various effects that need to be considered. This helps with informing and building the model itself, which is presented in Section \ref{sec:model}.

\subsection{Data sources and usage}

A large range of data sources has been used for the current research, as presented in Table \ref{tab:data_sources}. The year of reference was chosen to be 2014, which was the most recent available year of data across the different sources. 

A major input was airport financial and operational data sourced (through subscription) from FlightGlobal (London, UK). ATRS (Air Transport Research Society; USA and Canada) benchmarking study data were purchased, in addition, particularly for the provision of complementary data on airports' costs and revenues. At the time of analyses, only ATRS data for 2013 were available, and these selected data were used as a proxy for 2014, after checking their validity for this. Financial and operational data were compared with in-house, proprietary databases, with adjustments made as necessary. Data on airport ownership, and additional data on passenger numbers, were provided by Airports Council International (ACI) EUROPE (Brussels). European traffic data were sourced from EUROCONTROL's Demand Data Repository (DDR) with delay data primarily from the Central Office for Delays Analysis (EUROCONTROL, Brussels). Note that, whilst pure turnaround delay would ideally be used, as this reflects airport \textit{in situ} effects only, general (total) air traffic flow management departure delay was found to work as a statistically good proxy for this. Furthermore, we did not have access to clean, local (airport generated) air navigation service (ANS) delay data. Other in-house sources of data were used in addition to those listed, also drawing on the literature review.

Considering the wider context of operations in 2014, there were 1.7\% more flights per day in the EUROCONTROL statistical reference area, compared with 2013. The network delay situation remained stable compared to 2013, notwithstanding industrial action, a shifting jet stream and poor weather affecting various airports throughout the year, particularly during the winter months \citep{coda}. The average delay per delayed flight demonstrated a slight fall relative to 2013, and operational cancellations remained stable \textit{ibid}. The issue of industrial action, prevalent in 2014 in particular, was shown not to impact the model.

\begin{table}[htbp]
\begin{center}
%{\small
%\rowcolors{2}{white}{lighter_blue}
\begin{tabular}{C{0.3\textwidth}C{0.3\textwidth}C{0.3\textwidth}}
\hline
\textbf{Source} 			& \textbf{Typical Content} & \textbf{Use}\\
\hline
\hline
FlightGlobal & Number of flights, number of passengers, share of European flights & Cluster analysis, calibration\\
EUROCONTROL CODA 			& Delay per airport \& per cause of delay & Comparison with DDR delays\\
EUROCONTROL DDR 			& Full trajectories of aircraft for one month of data & Delay distribution, capacity fitting, airline traffic shares\\
ACI 			& Number of passengers (domestic, international, etc.) & Calibration purposes\\
Skytrax, etc	& Passenger satisfaction	& Cluster analysis\\
ATRS			& Financial data			& Cluster analysis, calibration\\
ATRS 			& Airport charges			& Comparison with aeronautical revenues per aircraft\\
Private communication, EUROCONTROL (2016)	& Maximum Take-Off Weight	& Cost of delay calibration\\
University of Westminster \citep{cost_delay} & Cost of delay				& Cost of delay calibration
\end{tabular}
%}
\caption{Data sources, content, and use.}
\label{tab:data_sources}
\end{center}
\end{table}

\section{Presentation of the model}
\label{sec:model}

\subsection{Description}

The model is based on several core ideas arising, in part, from the literature review. \correctionbis{It does not include every aspect presented in the literature, but rather tries to find the minimal modelling ingredients to capture the most important features, with sufficient data to be calibrated. In particular, demand management techniques have not been included in the model because they should only play a role after the main capacity (the infrastructure) has been decided. In fact, these demand management techniques affect the cost efficiency of the airport and as such are represented within its cost function, as described thereafter.}

First, it is necessary to select only the delay caused by a given airport, eliminating all delays triggered by other airports or other sources. A representative agent description is used, i.e. all the airlines are described by a single, average representation. The following mechanisms were selected for the model:

\begin{itemize}
\item Delay is created primarily by a shortage of capacity.
\item Delay has a direct cost impact on the airlines: passenger reaccomodation, crew costs etc.
\item Airlines try to avoid additional costs from delay and thus might decide to drop a route if the delay is too high.
\item Passenger choices are primarily driven by external, non-airport management choices (airport location, airline fare and service) and thus are not modelled here.
\item Airport revenues can be divided into two components: (i) aeronautical (depending directly on the number of flights); non-aeronautical (depending directly on the number of passengers).
\item Intra-day traffic patterns and distributions of delay should be taken into account due to the non-linearity of the cost of delay for airlines.
\end{itemize}
Based on these considerations, we build the model around the relationships presented below. Note that in terms of heterogeneity of traffic and delays, we use 1-hour time windows, from 0500 to 2200. For each of the time windows, we consider the average traffic, computed over one month of data to have a good estimation of the typical intra-day pattern. Moreover, within each time window we use a full distribution of delays. This distribution is thus different from one time window to another. Equations \ref{eq:delay}, \ref{eq:c_d}, \ref{eq:P_A}, and \ref{eq:r_A} presented below are applied independently of each of the time windows and the results are summed afterwards. \correctionbis{For the same reasons, the quantities involved in the equations are usually to be interpreted as `per hour'.}

% Based on these considerations, the following building blocks are considered for the model. First, regarding heterogeneity in terms of traffic and delay: hourly traffic is used and applied to time windows, that frame the various equations linking the parameters. Moreover, each time window has a different distribution of departure delays with, in particular, different means and variances. 

A given, constituent equation is defined for the relationship between the level of traffic and the delay generated. In order to do this, capacity is considered as an emergent property of the relationship between traffic and delay, more specifically, as the amount of traffic that the airport can handle before the delay increases. Based on the literature review \correctionbis{\citep{desart, wan} and on our own regressions (see calibration discussion)}, an exponential relationship is chosen between the number of departures \correctionbis{per hour} $T$ and the average delay at departure $\delay$ (\correctionbis{in minutes}):
\begin{equation}
\delay = 120 (\exp(T/\mathcal{C}) - cc),
\label{eq:delay}
\end{equation}
\correctionbis{where $cc$ is related to the delay generated when the traffic is very low, and $\mathcal{C}$ is the capacity. Hence, this equation can be considered as the definition of the capacity for an airport. It represents the }typical limit beyond which delay appears. \correctionbis{In particular, it is important to note that we do not assume a priori that the capacity is linked primarily to the runways or to the terminals, or that it increases linearly with the number of these infrastructures. The capacity as a whole is a complex interplay between numerous processes, which creates the delay.}

\correctionbis{Note also that considering that the delay within a time window is only dependent on the traffic within this time window is a simplification. In reality, the delay is also a function of the traffic within the previous time windows. This is not formally considered by the model, but is captured to some extent by the regression made during the calibration process.  Indeed, the direct effect of delay spilling over is the increase of delay in a time window where the traffic would theoretically be low enough to have a lower level of delay. This probably means that on average, low traffic time windows manifest a delay increase. As a consequence, this should be captured by the regression to some extent, with a greater weight applied to the low traffic periods.}

This delay has a cost \correctionbis{$c_d$} for the airlines, \correctionbis{and \citep{cost_delay} have shown that in general this} can be modelled as a quadratic function \correctionbis{of the delay duration}:

\begin{eqnarray}
c_d  & = 7.0\, \delta t + 0.18\, \delta t^2 + & (-6.0\, \delta t - 0.092\, \delta t^2) \sqrt{MTOW} \nonumber\\
	& & \mbox{if} \quad \delta t \ge 0,\nonumber\\
	& = 0. \quad & \mbox{otherwise}, \label{eq:c_d}
\end{eqnarray}
where $\delta t$ is the individual delay of a single flight \correctionbis{in minutes, $MTOW$ is the maximum take-off weight of the aircraft measured in metric tonnes and the cost is measured in euros}. This relationship has been obtained based on delay cost modelling by aircraft type and delay duration, undertaken from 2002, based on literature reviews, stakeholder inputs and industry consultations, the third phase of which was reported in 2015 \citep{cost_delay}. The equation above has only one parameter in addition to the delay itself\correctionbis{, which is the square-root of the maximum take-off weight} of the aircraft. It should be noted that this function is not linear, a) because of the quadratic term and, b) because `negative' delays (early departures) do not yield gains for the airlines. As a result, one cannot \correctionbis{directly replace} $\delta t$ by its average $\delay$ in this equation, but one needs to take into account the full distribution of delay. In particular, it is clear that even an airport with a null average delay has a non-null cost for its airlines. This point is crucial and is further studied in the calibration discussion of Section \ref{subsec:calibration}.

An increase in the cost of delay has a direct consequence of making flights less profitable for airlines. As a result, it is assumed that airlines tend to decrease their participation at an airport when this happens. For this, a logistic function is chosen, based on the cost of delay $c_d$ and a decision `smoothness' $s$ \correctionbis{(measured in the same units as $c_d$, i.e. euros)} as follows:
\begin{equation}
P_a = \frac{2}{1 + e^{c_d/s}},
\label{eq:P_A}
\end{equation}
where $P_a$ represents the probability of the airline actually operating the flight. This function is monotonically decreasing with the cost of delay. The parameter $s$ drives the choice of the airline, which immediately stops the operation of the flight as soon as the cost is greater than 0 if $s$ is small, but otherwise continues its operation even if the cost is non-null if $s$ is high. This models the fact that the airline does not base its decision only on one flight, but on its whole network, and is thus likely to even accept some loss if the flight brings some benefits elsewhere. \correctionbis{This function is clearly linked to the demand elasticity, but we chose this form of function because an earlier version of the model included some degree of risk aversion from the airline, naturally taken into account with this kind of function. This feature was removed because of the lack of distinct results with and without risk aversion and the difficulty to calibrate the risk aversion parameter.}

%This function is clearly linked to the demand elasticity, but this function is used instead of a linear relationship in order a) to capture non-linear effects and b) to introduce some heterogeneity in the airline business model at a later stage in the model.

\correctionbis{Note also that we did not consider the airport charges in the cost function of the airline. Indeed, some airlines are not particularly sensitive to airport charges, whereas others are. This depends on a number of factors such as the airline business model, length of haul, etc. Moreover, whilst some airports may be able to raise their charges, others will be constrained by being subject to formal economic regulation which may not allow this, or will have to consult and seek government approval for any increase. Therefore, due to the number of unknowns here, it was decided to keep airport charges constant in our analysis.}

The probability of operating the flight then fixes the actual traffic \correctionbis{(number of flights departing per hour)}, in the form of $T = P_a \beta$, where $\beta$ is the potential demand. However, in turn, this level of traffic changes the average delay, which changes the cost, the probability of operating the flight and so on. There is then the need to solve an implicit equation, which can be interpreted as an economic equilibrium with the mean delay playing the role of price (see \ref{annex:implicit}). This interpretation is important to bear in mind for the understanding of some of the results in Section \ref{sec:results}.

Once the traffic is known, the revenues of the airport are computed. It is assumed that the average number of passengers per flight $n_f$ is constant for all the flights at the airport, hence generating a linear relationship between the number of passengers and the number of flights. The revenues are divided into two components, as mentioned above:
\begin{itemize}
\item Aeronautical - linear in the number of flights, $T$.
\item Non-aeronautical - linear in the number of passengers, $n_f T$.
\end{itemize}
\correctionbis{Aeronautical revenues are indeed generally made up of a landing charge levied on the MTOW or MAW weight of the aircraft (which broadly correlates with passenger numbers) and a passenger charge levied per passenger. So both charges in effect are roughly based on passenger numbers. However when a constant average number of passengers per aircraft is assumed -- as it is the case here -- the weight will be constant and the revenue will increase linearly with the number of flights. Non-aeronautical revenues are very much driven by passenger numbers because if the airport operator provides commercial facilities themselves, it is normally the case that more passengers mean more spending. If the airport subcontracts out commercial facilities (which is more typical) the concessionaire will normally pay a fee based on their own revenue to the airport operator which again will be closely related to passenger numbers. }

\correctionbis{Ultimately, in} this framework both types of revenue are directly proportional to the traffic volume. Hence the total revenues have the form:

\begin{equation}
r_A = (P + n_f w) P_a \beta,
\label{eq:r_A}
\end{equation}
where $P$ represents the aeronautical revenues \correctionbis{in euros} per flight, $w$ are the non-aeronautical revenues per passenger\correctionbis{, and $n_f$ the average number of passengers per flight}. The former is considered fixed throughout this paper, since it \correctionbis{arises} mainly from airport charges, which are regulated in many countries and thus do not represent a variable of major adjustment for the airport, \correctionbis{as explained above}. The latter are considered fixed in this section and for the first results, but are relaxed in the last part of Section \ref{sec:results}, allowing for more complex behaviours from the passengers.

\correctionbis{Finally, we consider the \textbf{operating} cost $c_{inf}$ of having a capacity $C$ with a simple linear function}:
\begin{equation}
c_{inf} = \alpha  (\mathcal{C} - \mathcal{C}_{init}) + c_{init},
\label{eq:c_inf}
\end{equation}
where $\mathcal{C}_{init}$ is the current capacity of the airport, $(\mathcal{C} - \mathcal{C}_{init})$ represents, for instance, a planned increment of the capacity, \correctionbis{and $c_{init}$ represents the cost to operate the airport at capacity $C_{init}$. The costs are measured in euros per hour and the capacity in number of flights per hour.} The parameter $\alpha$ \correctionbis{-- in euros per flight per hour -- } is crucial here, because it represents the marginal operating cost of capacity, i.e. the cost of \correctionbis{operating} an extra unit of capacity. It should be noted that this form of the cost does not preclude its utilisation for discrete increments of the capacity, such as the construction of a new runway. The linear law can hold even in this case, because it only assumes that two runways would cost twice as much and yield approximatively twice the increase in capacity\footnote{\correction{This is obviously an approximation. Running a second runway is clearly not as expensive as running the first, for example because the control tower is already operating and would need relatively few enhancements. On the other hand, having two runways is not twice as efficient as having one, because of runway congestion and taxi times. Overall, a linear relationship seems to be reasonable as a first approximation. In particular, our point here is that the \textbf{operating} cost of running a given capacity is \textbf{not} a \textbf{highly non-linear} function of the capacity. This is in contrast with the \textbf{process} of extending capacity, which is achieved through discrete increments.}}. The only caveat is to consider $\mathcal{C}$ as a discrete variable instead of a continuous one, which is discussed in Section \ref{sec:results}. \correctionbis{We also emphasize that the quantity $c_{inf}$ is the \textbf{operating} cost for the airport, i.e. the cost of actually operating the airport on a day-to-day basis. In addition to labour, this includes contracted-out services, maintenance and repairs, administration, and other similar costs. As with the ATRS data, our definition does not include depreciation, although this does sometimes get included in airport accounts as operating costs. Other capital costs, such as the interest paid on new investment, are also not considered.}

Note also that the passengers are directly impacted by the delays. In particular, their desire to take a flight at a very congested airport might decrease, which could drive the profit of the airlines down \correctionbis{also}. This can be taken into account \correctionbis{through} the cost of delay of the airline, but is likely to be small in any case. More importantly, deriving passengers' preferences with regard to their time (`value of time' problem) and their decision-making process is a \correctionbis{distinct} area of research which is far beyond the scope of the present study. 
\\

\correctionbis{Note that in this model there is no profit maximisation for the airport. Instead, we aim at deriving its operating profit based on different parameters in order to potentially help decision makers regarding capacity expansion.}

\correctionbis{Note also that the model is fully deterministic and does not take into account any kind of uncertainty a priori. In fact, in the calibration section we include the uncertainties of the delay, which have a strong effect on the cost of delay of the airlines. Most of the other quantities are fully deterministic however, mainly due to the lack of data for calibration. The agents also do not exhibit any kind of risk aversion, as previously emphasised, because of the difficulty of calibrating risk aversion and the overall lack of information concerning this point.}
\\

The five constituent equations \ref{eq:delay}, \ref{eq:c_d}, \ref{eq:P_A}, \ref{eq:r_A}, and \ref{eq:c_inf} form the backbone of the model. The parameters in these equations \correctionbis{are summarised in Table \ref{table:calibration}} and can be estimated from data as described \correctionbis{thereafter}. 

\begin{table}[htbp]
\begin{center}
\begin{tabular}{cccc}
\hline
Name		& Description 		& \begin{tabular}{c}Type of\\ parameter\end{tabular} & Unit \\
\hline
\hline
$MTOW$					& Max. take-off weight		& DC					& \correctionbis{metric tonnes}\\
$n_f$					& \begin{tabular}{c}Average number of\\passengers per flight\end{tabular}	& DC & \correctionbis{pax per flight}\\
$P$						& Airport charges			& DC					& \correctionbis{euros per flight}\\
$\mathcal{C}_{init}$	& (Departure) capacity		& DC					& \correctionbis{flights per hour}\\
$cc$					& Delay \correctionbis{offset} at zero traffic		& DC					& \correctionbis{minutes}\\
$v$						& Value of time 			& DC					& \correctionbis{euros per minute}\\
$T$ 					& Distribution of traffic 	& DC					& \correctionbis{flights per hour}\\
$w$ 					& Average revenue per passenger 	& DC			& \correctionbis{euros per pax}\\
$c_{init}$				& Total initial cost		& DC					& \correctionbis{euros per hour}\\
$\beta$					& Traffic multiplier (demand)	& PC				& \correctionbis{flights per hour}\\
$\alpha$				& Marginal cost of capacity & FP					& \begin{tabular}{c}euros per\\ flight per hour\end{tabular}\\
$s$						& Smoothness				& FP					& \correctionbis{euros}\\
\hline
\end{tabular}
\caption{List of parameters of the model, with their types related to calibration. DC: direct calibration, FP: free parameter, PC: post-calibrated. \correctionbis{See Section \ref{subsec:calibration}}.}
\label{table:calibration}
\end{center}
\end{table}

\subsection{Calibration}
\label{subsec:calibration}

The calibration of the model deploys three steps:
\begin{itemize}
\item The direct calibration, whereby a parameter of the model is directly related to a value which can be extracted from the available data.
\item The functional relationship calibration, whereby a function between two observables is matched to the data, sometimes using a regression to fix some parameters.
\item The post-calibration, whereby a parameter of the model is unobservable. In this case, the values of the parameter are swept, measuring an output of the model and trying to match it to an observable target from data.
\end{itemize}

\subsubsection{Direct calibration}

The first step allows estimates of different parameters of the model such as:
\begin{itemize}
\item The average number of passengers per flight $n_f$ is given by the ratio of the number of flights and the number of passengers. 
\item The (average) aeronautical revenues per flight $P$ are given by the total aeronautical revenues divided by the number of flights.
\item The (average) non-aeronautical revenues per passenger $w$ are given by the total non-aeronautical revenues divided by the number of passengers.
\item The distribution of traffic $\{T\}$ through the day is fixed by averaging one month of data, splitting the day into \correctionbis{1-hour windows}.
\item The average square-root of the maximum take-off weight $MTOW$, based on the individual weights of the aircraft operated by the airlines at the airport.
\end{itemize} 

\subsubsection{Functional relationships calibration}

The second step of the calibration is to build functional relationships between some variables of the model through regression. A particular case is that of the relationship between the average delay and the level of traffic. For this, a least-square exponential fit of the delay against the number of departures per hour over one month of data was performed. This yields the value of $cc$, which is linked to the delay at low traffic (usually negative), and the capacity $\mathcal{C}$. It should be noted that performing this fit, or a linear fit, usually yields similar results in terms of goodness of fit (with a $R^2$ between 0.6 and 0.9 for most of the airports), thus challenging the usual use of an exponential function from the literature.

A second important relationship to be calibrated is the cost of delay. Whilst the average of $\sqrt{MTOW}$ can be easily directly fixed from the data, account needs to be taken of the distribution of delay in order to compute the average cost of delay. This is done in three steps. First, for a given airport and for each hour of the day, the empirical distribution of delay is built, and then a fit with a log-normal distribution is performed. The reason to use a fit rather than the empirical distribution is to allow for easily \correctionbis{adjusting} the parameter of this distribution afterwards, in particular its variance, linked to the predictability of the departure times. The specific choice of a log-normal distribution over other distributions is based on a) its simplicity in terms of parameters and, b) its fundamentally asymmetric shape, with a few rare events at very high delays.

\begin{figure}[htbp]
\begin{center}
\includegraphics[width=0.9\textwidth]{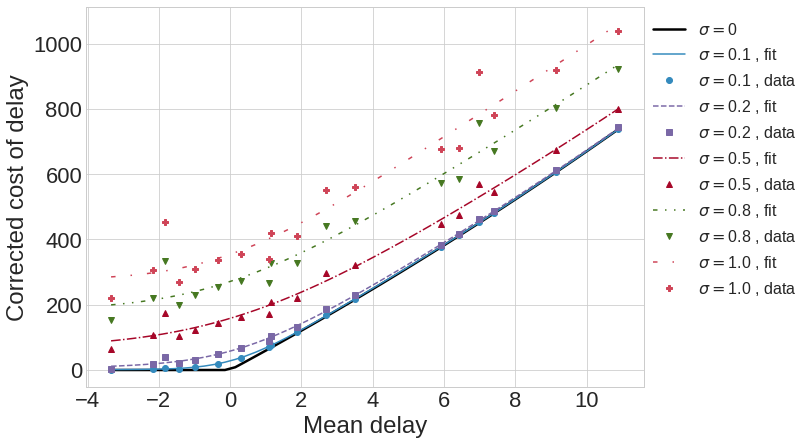}
\caption{Evolution of the expected cost of delay based on a log-normal distribution of the delays. The black line represents the cost when the variances of the distribution tend to zero. The coloured points are the expected values for a given airport at different times of the day (with different mean delay). Different colours represent different values of the standard deviation of the distributions. \correctionbis{The standard deviations are normalised, so} `$\sigma=1$' \correctionbis{represents the standard deviation found originally in the data, `$\sigma=0.5$' half of the standard deviation found in the data, and so on}. Finally, the solid coloured lines are obtained via regression using a quite complex function, see \ref{annex:cost_delay}.}
\label{fig:correction}
\end{center}
\end{figure}

With the distribution for each hour, the expected value of the cost of delay is simply obtained, using:
\begin{equation}
\overline{c_d} = \int_0^\infty c_d( \delta t ) p(\delta t)\, d (\delta t),
\label{eq:expected}
\end{equation}
where $p(\delta t)$ is the probability of having the delay $\delta t$, based on the log-normal distribution described previously. Since for each hour of the day there is a different value of the mean delay $\delay$, a plot of the equivalent of Equation \ref{eq:c_d} with the expected cost against the mean value can be made and compared with the cost of the average delay (replacing $\delta t$ by $\delay$ in Equation \ref{eq:c_d}). This plot is shown in Figure \ref{fig:correction} for a particular airport in the database, where it can clearly be seen that the average cost of delay is significantly different from the cost of the average delay.

The final step concerning the cost of delay is to perform a fit in order to use it as a continuous variable in the model. This is done by using a complex function, as explained in \ref{annex:cost_delay}. The result of this regression is shown in Figure \ref{fig:correction}, with solid lines. The regressions are robust for most of the airports considered in this paper ($R^2>0.95$). 

\subsubsection{Post-Calibration}

The last step of the calibration process is to sweep the unobservable parameter $\beta$ in order to match an output of the model with its value in the data. For this, the total number of flights operated at the airport within each one-hour window is used. Increasing $\beta$, the model will slowly increase the total number of flights in output and this is stopped when this value matches the one extracted from the data for this time.

\subsubsection{Summary of calibration}

In summary, the calibration process includes the following steps:
\begin{itemize}
\item Maximum take-off weights $MTOW$ are included in the cost-delay relationship.
\item Average number of passengers per flight $n_f$, aeronautical revenues per flight $P$ and non-aeronautical revenues per passenger $w$, value of time $v$, total initial cost $c_{init}$, and distribution of traffic $\{T\}$ through the day are taken directly from data.
\item Fitting parameters $cc$ and $\mathcal{C}_{init}$ (the latter being the capacity) for delay-traffic load relationships are set.
\item The cost of delay relationship is corrected based on intra-hour log-normal fitting distributions of delays.
\item A demand factor $\beta$ is post-calibrated by matching the number of flights with the data.
\end{itemize}
Note that the ``total initial cost $c_{init}$'' represents the total current costs of the airport, i.e. the costs for providing the current capacity.

Finally, there are two parameters remaining, the smoothness of the airline decision $s$ and the marginal cost of capacity $\alpha$, which is the cost of operating one extra unit of capacity. The latter could be estimated, for example, by considering that the primary mission of an airport is to deliver capacity for flights, and thus that all its costs are related somehow to this mission. Hence, dividing the current capacity by the total costs would give the marginal cost of capacity. This, however, should only be considered as a rough estimation, and $\alpha$ is considered as a free variable in the following. 

The smoothness $s$ is thus the last free parameter of the model. It represents the sensitivity of the airline to the cost of delay, which is very hard to estimate because of the lack of detailed airline data. It is worth noting, however, that:
\begin{itemize}
\item A basic sensitivity analysis (see \ref{annex:sensitivity}) shows that the results of the model do not depend strongly on the value of $s$.
\item The parameter is actually not totally free, but is constrained at low values. This is because a low elasticity cannot fulfil demand requirements.
\end{itemize}

Table \ref{table:calibration}  \correctionbis{presented a summary of how the parameters are calibrated}.

\section{Results}
\label{sec:results}

In this section we present the results obtained with the model. This begins with some results with the model calibrated on a large European hub. Then the impact of different parameters on the results is shown, before comparing different airports. Finally, some results obtained with more \correctionbis{a} complex behaviour of the passengers are presented. 

\subsection{Profit evolution for a large hub}
\label{net_income}

\begin{figure}[htbp]
\begin{center}
\includegraphics[width=0.75\textwidth]{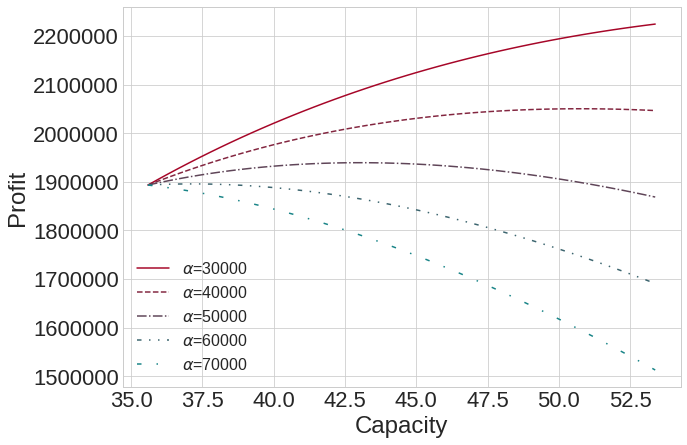}
\caption{\correctionbis{Daily} profit of the airport as a function of capacity, for different values of the marginal operating cost $\alpha$.}
\label{fig:net_income}
\end{center}

\end{figure}

Following the procedure described previously, firstly the model is calibrated on a large European hub. In order to see if a potential increase in the capacity would be profitable for this airport, the plot in Figure \ref{fig:net_income} presents the \correctionbis{operating} profit of the airport as a function of the capacity and the marginal \correctionbis{operating} cost $\alpha$. The figure shows that for high values of $\alpha$, the profit decreases monotonically with the capacity, because capacity is very expensive in this case. When $\alpha$ has an intermediate value, there exists a unique maximum in the profit, whereas when $\alpha$ is low, the profit increases monotonically because the capacity is essentially free.

The presence of an optimum is important for the airport: it means that the airport could potentially increase its revenue by increasing its capacity. As already noted, an airport cannot usually increase its capacity continuously, but rather by discrete increments, e.g. by building a new runway. The graph shown in Figure \ref{fig:net_income} shows the possibility of assessing the profitability of the increment, by comparing the expected profit with the extra capacity, to the profit with the current capacity. 

\correctionbis{It is also interesting to find the average delay that corresponds to the optimal state -- the maximum profit. If one takes the marginal cost of capacity $\alpha$ to be 60 000 euros per hour, comparable to the current cost of the airport of running its current capacity, one finds that the optimal average delay is around 9.5 minutes, slightly below the current delay of 9.6 minutes for this airport. The gain in punctuality in this case is thus small both for the airport and for the passengers.}

\subsection{Effect of the average number of passengers per flight on the optimal state}
\label{load_factor}

The presence of an optimal capacity for the airport is important, but it needs to be assessed whether this could be affected by different parameters. The first is the average number of passengers per flight at the airport $n_f$. This is motivated by previous research that reports that an increase in the average number of passengers per flight has been used by airlines at congested airports as a relatively cheap way of increasing their capacity \citep{berster}. It should be noted that, in principle, an increase in the average number of passengers per flight has no impact on runway capacity but can affect terminal capacity. However, it seems clear from the literature review that the current bottleneck is the runway and not the terminal, at least for highly congested airports \citep{gelhausen,berster}.

\begin{figure}[htbp]
\begin{center}
\includegraphics[width=0.75\textwidth]{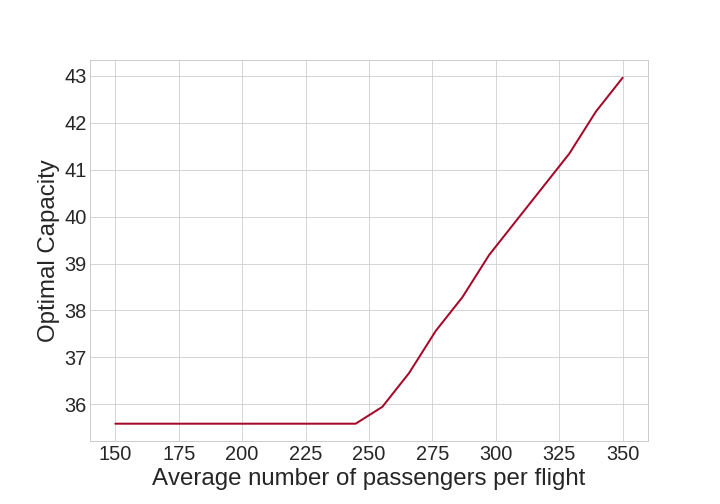}
\caption{Evolution of the optimal \correctionbis{hourly capacity} as a function of the average number of passengers per flight.}
\label{fig:load_factor}
\end{center}
\end{figure}

\correctionbis{In order to investigate this,} the model is calibrated on the same airport as above and then the average number of passengers per flight is changed. The capacity is also swept to detect the position of the optimum as a function of the average number of passengers per flight. The marginal cost of capacity $\alpha$ is fixed \correctionbis{once again at} 60 000 euros per hour. Figure \ref{fig:load_factor} displays the results of the procedure. \correctionbis{In this plot, we have capped the optimal capacity such that it does not fall below the current capacity.} As a result, increasing the average number of passengers per flight at first does not change the optimal point, which is the current capacity. Going further, the position of the optimum then increases linearly with the average number of passengers per flight. This \correctionbis{happens because} a higher average number of passengers per flight will create a higher yield for the airport when attracting new flights, which pushes the optimal capacity upwards. 

This simple linear relationship could be easily used as a rule of thumb for airports. For instance, instead of considering an increase in capacity to decrease the delay by X\%, the airports could try to incentivise airlines to increase their average number of passengers per flight by Y\%. This simple relationship can also be used to roughly predict when the average number of passengers per flight at the airport will increase, based on the congestion at the airport and what its optimal capacity would be.  

\subsection{Effect of the predictability on the optimal state}
\label{predictability}

Of further interest is the effect of predictability. Many stakeholders, including passengers and airlines, use significant buffers because of the uncertainty in the system, which leads to longer travel times. Once again, using the calibrated model for the same airport, the effect of predictability by varying the distribution of delays is studied. As previously described, based on real delay data, a log-normal fit is used to simulate the delay and compute its cost. In this experiment, the variance of these distributions (for each one-hour time window of the day) is decreased, keeping the means constant. This simulates a situation where the predictability is increased while the punctuality (mean delay) is fixed. More specifically, the standard deviation of all the distributions during the day is reduced by the same factor. Once again, $\alpha$ is fixed and the capacity swept to detect the optimal value. 

\begin{figure}[htbp]
\begin{center}
\includegraphics[width=0.49\textwidth]{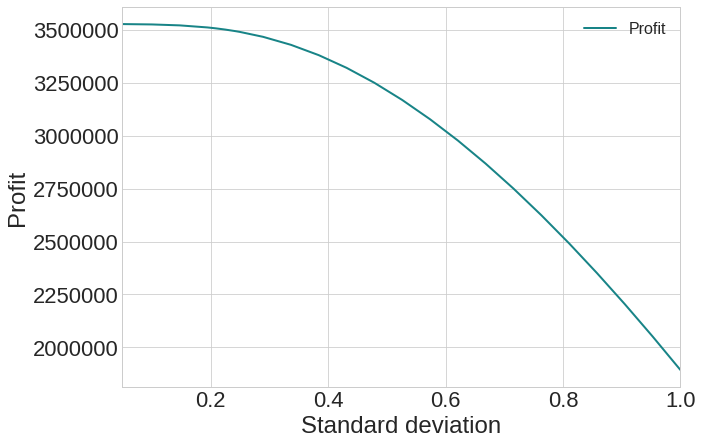}
\includegraphics[width=0.49\textwidth]{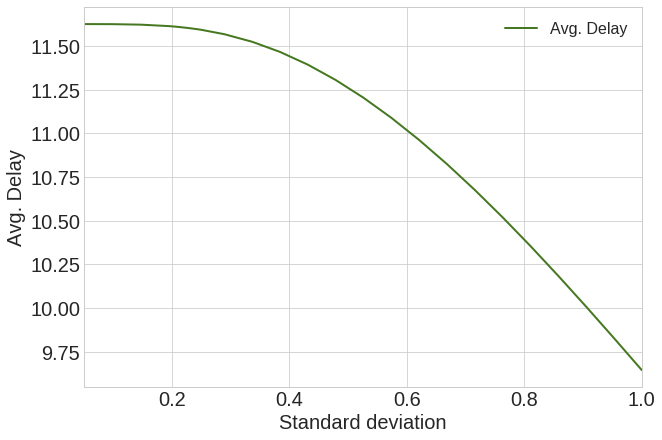}
\caption{Left panel: \correctionbis{daily} profit as a function of the standard deviation of the distribution of delay. Right panel: average delay \correctionbis{in minutes} against the standard deviation of the distribution of delay. The standard deviation is actually measured with respect to its initial value. Hence, a standard deviation of 1 represents the state where the initial predictability is used and 0 represents the perfectly predictable case.}
\label{fig:predictability}
\end{center}
\end{figure}

To understand the impact of predictability, the left panel of Figure \ref{fig:predictability} shows the evolution of the profit of the airport for a fixed capacity against the reduction of the standard deviation. As expected, profit grows as predictability is increased (from right to left on the graph). %This comes directly from the fact that the cost of delay gets smaller with higher predictability, hence operating the airline is more profitable and they are attracted to the airport. 
However, there is a striking side effect, which is that the average delay at the airport actually increases with the predictability, as displayed in the right panel of Figure \ref{fig:predictability}. 
%This comes simply from the fact that there are more airlines attracted to the airport, and thus the airport is more congested. 
In other words, there seems to be a trade-off between predictability and punctuality at an airport.

In order to understand this mechanism, reference is made to the resolution of the implicit equation explained in Section \ref{sec:model} and \ref{annex:implicit}. The direct effect of the reduction of uncertainty is the decrease of the correction term applied to the cost of delay as computed by Equation \ref{eq:expected}, i.e. a direct reduction of the cost of delay. As a consequence, for a given mean delay, the airline is more willing to operate a flight, which drives the demand function of Figure \ref{fig:implicit} up. Since the supply is unchanged, this means that the delay at equilibrium is higher than before, which explains the behaviour of Figure \ref{fig:predictability}, when the uncertainty starts to decrease. Conversely, it also explains the increase in the profit of the airport, since airlines are more willing to operate at the airport at no extra capacity cost\footnote{In reality, this increase in predictability is likely to be the consequence of the adoption of some technology, which has a price. This price, which is likely to be shared among several stakeholders, is not taken into account here.}. This effect is counter-intuitive but is \correctionbis{equivalent} to an increase in price due to easier access to a market of commodities\footnote{Average delay and its variance are actually correlated. This does not change the conclusion of the model, in the sense that the impact of the variance is as described. This neglects behavioural feedback and additional effects from concomitant changes in punctuality introduced by the new technology/procedure.}.

\begin{figure}[htbp]
\begin{center}
\includegraphics[width=0.75\textwidth]{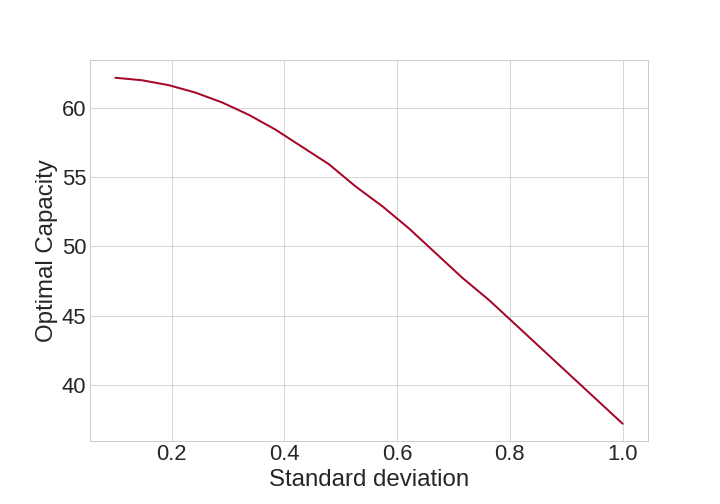}
\caption{Evolution of the optimal \correctionbis{hourly} capacity as a function of the normalised standard deviation of the distribution of delays.}
\label{fig:predictability_opt}
\end{center}
\end{figure}

In any case, the position of the optimal capacity for the airport is likely to be modified by the predictability. This is indeed the situation, as shown in Figure \ref{fig:predictability_opt}. When the predictability increases, the optimal capacity increases too, essentially because the airport is able to manage more flight\correctionbis{s} with the same level of delay. It should be noted that this effect is linear at first, but saturates when reaching very small deviation\correctionbis{s}. This region is probably unrealistic in any case, because the mean \correctionbis{(arrival)} delay would probably drop when the predictability decreases so much. \correctionbis{High systematic delay, driven by low predictability, would be predicted by the airlines and off-set through increased buffers and earlier departures, for example, thus reducing the arrival delay.} 
\\

Finally, note that the effect of predictability is far from being negligible. According to the model, even an increase of 10\% in predictability could lead to an increase of 16\% in profit, with only a 4\% increase in the mean delay (less than a minute). Of course, the (\correctionbis{operating}) cost of the improvement of the predictability is not taken into account here, and could drastically change the picture.

\subsection{Comparison between airports}

%One of the objectives of the project was to build a unified model able to forecast the value of extra capacity at different airports.
So far, the results of the model for one airport have been investigated. The differences between airports are now considered as it is clear that different airports can sustain different costs, in particular regarding the \correctionbis{operating} costs related to extra capacity. To study this point, an increment of one unit of capacity is assumed for all airports in the database and the value of $\alpha$ is found where the profit of the airport would be the same as with the original capacity. This value of $\alpha$ indicates the maximum \correctionbis{operating} cost for which an extra unit of capacity becomes profitable for the airport. 

\begin{figure}[htbp]
\begin{center}
\includegraphics[width=0.75\textwidth]{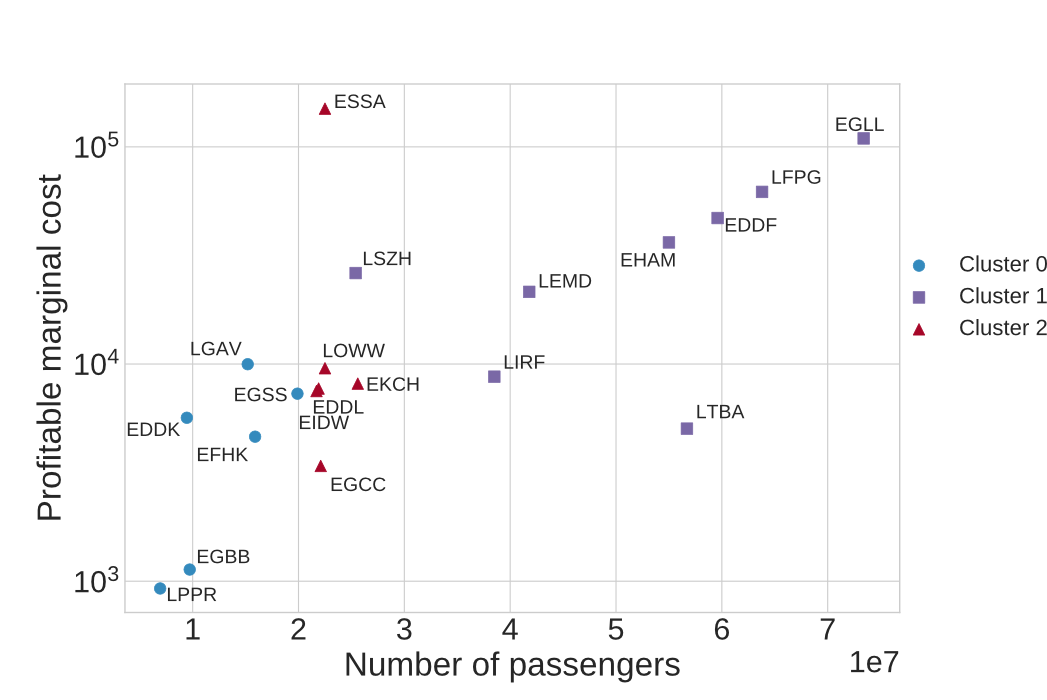}
\caption{Comparison for different airports of the maximum marginal cost for which an increase of one unit of capacity is profitable \correctionbis{against their yearly number of passengers}. The colour refers to different types of airports, as derived in \citep{sid_paper}, which roughly corresponds to large hub/small hubs/non-hub airports.}
\label{fig:alpha}
\end{center}
\end{figure}

The results are displayed in Figure \ref{fig:alpha}. The first conclusion is that different airports have very different levels of profitability, from around 1 000 euros per hour to more than 100 000 euros. Clearly, larger airports can more easily sustain an increment in capacity, simply because of their different \correctionbis{operating} expenses and revenues. When the profitable level is compared to the total \correctionbis{operating} costs at the airport, the dependence on the total number of passengers disappears, as shown in \ref{annex:comparison}.

It should be noted, however, that this dependence with size is far from perfect. In particular, some large airports (such as Istanbul Atat\"{u}rk airport) have a smaller profitable level than much smaller airports, such as Hamburg. This is also expected since different airports should have different needs in terms of capacity. In particular, the profitable level $\alpha$ is expected to be higher for airports which are already highly congested. Furthermore, national, or even regional, characteristics have to be taken into account, since the \correctionbis{operating} costs depend on the types of airport, the economic development of the country, and so on. The figure, however, shows a high-level picture which can be used to compare concisely and consistently the states of different airports. 

\subsection{Exploratory results}
\label{exploratory}

In this section the assumption of constant non-aeronautical revenues per passenger as applied previously is relaxed. Since there are no \correctionbis{public} data on the precise behaviour of passengers at an airport, the model cannot be completely calibrated. Therefore, this is only an illustration of the potential impact of different mechanisms. From the literature, two possible mechanisms emerge. Passengers may spend more if they:
\begin{itemize}
\item Have a longer airport dwell (waiting) time.
\item Are more satisfied. 
\end{itemize}
It is interesting to note that these two effects work in opposite directions when delays are present. Delays increase the waiting time, leading to potentially longer shopping time, but they typically decrease the overall satisfaction of the passengers, which would lead to a lower quality of the shopping time. It is difficult to assess if these two effects have the same magnitude in reality, for instance by cancelling each other out.

In order to illustrate these effects, `more shopping time' and `better shopping time' are assumed to have effects on different time scales. More specifically, it is assumed that small delays are relatively neutral from the satisfaction point of view, but that higher delays have a relatively larger effect. On the other hand, it is assumed that the passengers have a constant probability of spending a fixed amount of money per unit of time. These two assumptions result in the following functional forms:
\begin{equation}
w(\delta t) = w_{init} + w_{shop}(\delta t) + w_{sat}(\delta t),
\end{equation}
where:
\begin{equation}
w_{shop}(\delta t) = t_e \frac{\delta t -\delta t_{init}}{120} w_{init},
\end{equation}
and:
\begin{equation}
w_{sat}(\delta t) =  
\begin{cases}
s_e \left(\frac{\delta t - \delta t_{init}}{120}\right)^2 w_{init} & \mbox{if} \quad \delta t < \delta t_{init}\\
-s_e \left(\frac{\delta t - \delta t_{init}}{120}\right)^2 w_{init} & \mbox{otherwise}.
\end{cases}
\label{eq:w_new}
\end{equation}
By tuning the parameters $s_e$ and $t_e$, we are able to create non-trivial patterns for $w$. As already stated, the absolute values of these parameters are of relatively little importance. However, the model is kept self-consistent by setting $w$ to the constant value $w_{init}$ used in the previous version of the model when $\delta t = \delta t_{init}$, the average delay at the airport.

\begin{figure}[htbp]
\begin{center}
\includegraphics[width=0.49\textwidth]{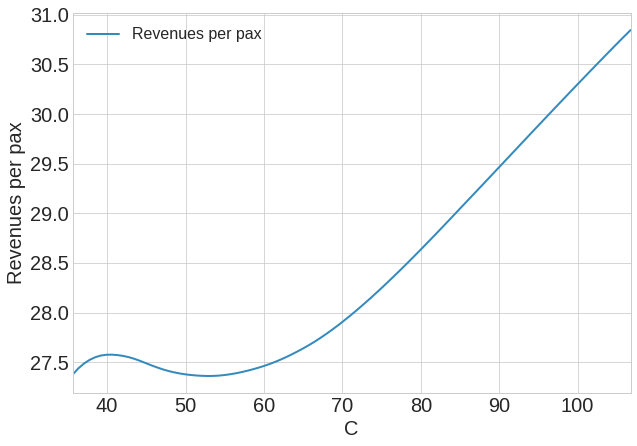}
\includegraphics[width=0.49\textwidth]{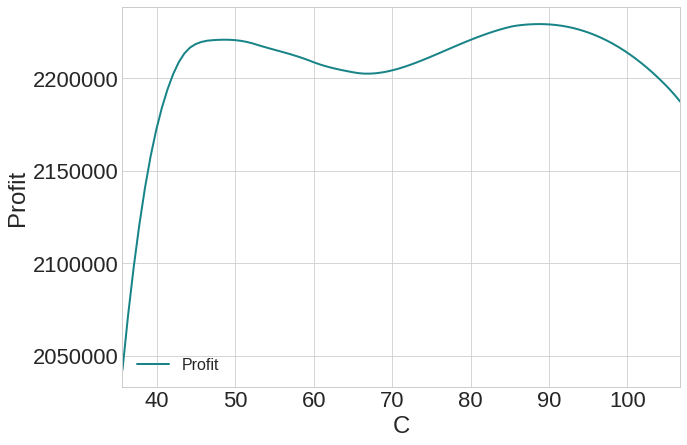}
\caption{Revenues per passenger (left) and total \correctionbis{daily} profit (right) for the airport as a function of the capacity when non-constant non-aeronautical revenues per passenger are assumed.}
\label{fig:w_new}
\end{center}
\end{figure}

\correctionbis{Combining this function with} Equation \ref{eq:r_A}, the model calibrated on the large European hub as in sections \ref{net_income}, \ref{load_factor}, and \ref{predictability}, is again used. The results for the revenues per passenger and the profit of the airport are presented in Figure \ref{fig:w_new}. As expected, the (total) revenue per passenger for the airport is no longer constant, but first decreases with the capacity, before increasing again. This shape now has a subtle interplay with the increased demand from $P_A$ (not shown here) to produce the shape of the profit curve on the right. 

This curve does not have a unique maximum as was the case previously. The presence of two maxima for the profit could have different consequences. Indeed, on the one hand, considering the air traffic management system as a stochastic system, where the airport seeks to maximise its profit, it could very well be that the airport would be `trapped' in a local maximum, instead of reaching the global maximum. Reasons for this could be economic risk-aversion, where an optimal choice, in principle, is discarded in favour of a lower-risk one, or simply the difficulty of raising investment capital, or overcoming regulatory constraints. On the other hand, the presence of a local maximum could actually be (temporarily) beneficial in some respects, where the airport waits for more investment or a better future solution. Regardless of the characteristics of the profit landscape, an important point is that the airport could be de-incentivised from investing in capacity infrastructures because delay could be beneficial to it, to some extent. 
\correctionbis{Indeed, in this case, the profits for the airport are close to each other at the two optima, but the gains for the passengers are quite different. Whereas the first one corresponds to an average delay of approximately 8.7 minutes, the global maximum reaches approximately 7.8 minutes, to be compared with the initial value, of approximately 9.6 minutes.}
This is an issue that regulators could tackle with the right incentive \correctionbis{or performance} scheme.

\section{Conclusions, assumptions, and future work}
\subsection{Conclusions}
In this article we have presented a simple model of an airport capturing the trade-off between an increase in capacity and its associated costs. Indeed, an airport operating close to its operational capacity is very likely to produce flight delays. These delays represent a direct or indirect cost for the airlines, which decreases the attractiveness of the airport as a business environment. This can decrease traffic demand, which represents an indirect cost of congestion for the airport. The balance between the \correctionbis{operating cost of providing} extra capacity and the shortfall due to congestion leads to the presence of an optimal capacity.

A simple deterministic model based on several functional relationships has been designed to capture this effect, and its magnitude, with the help of numerous sources of data. Among them, taking into account the full distribution of delay instead of the simple average, has proven very important to compute exactly the cost of delay for the airlines.

We have also shown that the position of the optimal capacity depends on several parameters. Among them, the average number of passengers per flight and the predictability of the flights, are the most important. Indeed, the average number of passengers per flight is currently regarded as a relatively cheap way of increasing the effective capacity of an airport, and it is important to study to what extent this can continue in the future for different airports. Even more important, unpredictability is supposed to decrease significantly in future, thanks in particular to various technologies envisioned by SESAR. It is also important to realise that an increase in predictability can produce, in principle, a sizeable decrease in the cost of delay for the airlines. Moreover, such an increase in predictability may lead \textit{in fine} to a degradation of punctuality, since the average congestion will increase, as the airport is more attractive. Note, however, that we do not explicitly model delay formation. Complex relationships between average delay (punctuality) and its variance (unpredictability) can arise in practice. A simple queuing model could, for example, be integrated into the model to reflect this.

We also showed that the airport may \correctionbis{unintentionally thwart an ultimate key goal}, considering that delay can increase the non-aeronautical revenues of the airport -- up to a certain point. This can decrease the incentive of the airport to increase its capacity, trapping it into some intermediate state where neither its revenues nor the passenger/airline satisfaction are maximal. This could be tackled by the right incentive scheme. We are, however, unable to draw conclusions regarding the presence of this effect in reality, due to the lack of data.

\subsection{Assumptions}

The model we present in this article makes several simplifications and hypotheses. Concerning the airports, most do not have the simple objective of maximising their profit. Indeed, there is a whole spectrum of airport governance in Europe, ranging from fully private (for which we could assume that they are indeed profit maximisers) to fully public (where considerations other than profit are taken into account). However, the model presented here does not assume that the airport maximises profit. Since the model is able to compute the profit in different situations, the information could be used in a wider cost-benefit approach, e.g. balancing optimal capacity and additional local noise, or the quality of service. 

We use the important concept of `capacity' for the airport. Usually, capacity is viewed as a hard constraint which cannot be exceeded. We argue that this vision is insufficient because, even if such a hard constraint exists, capacity has different consequences far before this constraint becomes limiting. For instance, it is clear that delays at an airport appear even before the declared capacity is reached, and grow rapidly with the traffic close to this limit. As a consequence, our view is that the capacity should rather be viewed as elastic. The consequences of having finite capacity are many, but one of the most important is the generation of departure (and arrival) delay at the airport. 

As a consequence, we define airport capacity in the model as arising from a purely phenomenological law between delay and traffic, computing its value with a regression on the appropriate data. In particular, we do not use the declared capacity of the airport, and we do not assume the source of the delay itself. Indeed, it is known \citep{gilbo, wan} that capacity can be broken down into terminal and runway capacities, but we do not need this distinction here since capacity is an emergent property of the airport performance data. The exact relationship between delay and traffic can be very complicated. In the model, average delay is associated with average traffic, using one-hour windows for the averages. A first problem with this choice is that the intra-hour variance of departure times could play a significant role. This could be fixed by reducing the time window, but at the risk of losing the more systemic effects whereby flights have a broader impact on airport congestion. This leads to the second issue, as there could be correlations between time windows if the latter are too small. Massive congestion in the morning would have consequences into the afternoon operations, for example. To capture this effect, one would need to make a regression with lagged variables with more coefficients than capacity alone, which is out of scope for the present article but planned for a future study. 

Related to capacity, we also need the model to have an estimation of the cost thereof. Due to the heterogeneity of the situation of airports, it is difficult to devise a general law. However, we argue that a linear law is our best estimate. Indeed, it is known that some airports display economies of scale \citep{bottasso, main, martin2011, voltes}, which means that their capacity should increase faster than their cost. On the other hand, incrementing capacity at an airport is not always easy, especially for large airports, and does not yield the same benefits as initial increments. Indeed, having two runways, for instance, does not provide double the capacity of one. As a consequence, we use a linear law in the model, the coefficients being estimated as explained in Section \ref{subsec:calibration}. 

Another issue is that demand at airports changes over time, and cannot be perfectly predicted. Airports have to consider medium- and long-term changes in traffic -- some of them easily predictable (e.g. seasons), others less predictable (e.g. economic crises). In the model, we assume that a reliable forecast for demand is available, and in practice one should use the best prediction of the traffic for a given future in order to have the best estimate of the optimal capacity. More importantly, it is easy to use the model in different traffic conditions, compare the levels of profit, and make an informed decision on whether the airport should expand its capacity or not. The uncertainties in the system (demand, other costs) are easy to take into account too, and the model can simply compute the profit and optimal capacity in different scenarios. The likelihood of having a given scenario must be computed independently, and an estimation of the expected profit/optimal capacity can thus be obtained. Regarding uncertainties, no risk aversion is included in the model, since it is intended to be a tool to assess the financial situation of airports, and not how they would react to a given situation -- which can be irrational to some extent, including some degree of risk aversion. 

We also consider that the aeronautical charges are fixed for the airport. This is a simplification arising from the fact that airports may be very differently regulated. Some of them are free to set these charges, but others have their charges controlled by a regulator in a number of different ways, such as with a price cap. As a consequence, we decided to keep them fixed. A more realistic model would allow a double optimisation with regards to charges and capacity for the most liberalised airports, which is also planned for future work.  

Another simplification of the model is that the number of passengers per aircraft is assumed to be constant. Some authors \citep{berster} have reported that the major airports in Europe seem to undergo a transition where the average aircraft gets larger and accommodates more passengers, which is a simple way (for the airport) to increase passenger throughput, but other authors report otherwise \citep{evans}. The change in the number of passengers per aircraft is  important, and is indeed reflected in the model by the parameter $n_f$, which can also be tuned to match various predictions. Moreover, we believe that the heterogeneity of this number among aircraft has a small impact and that an average value is sufficient at this stage. 

\subsection{Future work}
 
\correctionbis{This model should be seen as a first step towards a more detailed description of the costs and benefits of enlarging the capacity at different airports, to serve as a guide for different decision makers. In particular, the model should also take into account the changes in the demand landscape since the construction of a new runway, for instance, will be finished at a point in time where demand will be different from the current situation. Provided that good demand forecasts exist, they can be easily incorporated into the model, for example by adjusting the parameter $\beta$ to increase the overall demand, or by changing the daily pattern during the calibration phase.} 

Further developments of the model are planned through the use of other sources of data. In particular, it is important to take into account the reaction of passengers to delay, since they are the ultimate consumers. A step in this direction will be made by including a utility function for passengers linked to their value of time. Another planned development is the use of better cost functions for different types of airline. The data needed for this are highly sensitive, but we have already made several advances in this direction. We also plan to further our research into  network effects and how the delay created at a given airport propagates to others, thus decreasing the willingness of the latter to improve its facilities.

\section*{Acknowledgements}

This study was commissioned by the Airport Research Unit of EUROCONTROL to the University of Westminster and Innaxis, as part of its contribution to SESAR Operational Focus Area (OFA) 05.01.01 entitled `Airport Operations Management', in relation to the development of the economics and trade-off aspects of the APOC concept. We are most grateful to ACI for consultations on its Airport Service Quality data. Data on airport ownership, and additional data on passenger numbers, were kindly provided by ACI EUROPE. In particular, we thank Denis Huet, of EUROCONTROL, for his contributions and support during the course of this work. We are also most grateful to the reviewers for very helpful feedback in helping us to improve this paper.

\section*{Bibliography}
\bibliography{bibliography}{}

\newpage

\appendix

\section{Correction for the cost of delay}
\label{annex:cost_delay}

In this \correctionbis{appendix} the last stage of the correction of the cost of delay function as described in  \ref{subsec:calibration} is briefly discussed. For each one-hour time window, the expected cost of delay is computed, taking into account the probability of a given delay, based on European data. The results are shown in Figure \ref{fig:correction}, in the form of 18 points for each value of the normalised variability, corresponding to the 18 time windows considered within the model. Since each time window also has a different value for the mean delay, the result is a functional relationship between the expected cost of delay as a function of the mean delay. In order to be usable within the model, where the mean delay is a continuous variable, a suitable fitted function was sought. The aim here is to have a reasonably good approximation of the individual points rather that a deep understanding of the underlying mechanism behind the relationship. 

\begin{figure}[htbp]
\begin{center}
\includegraphics[width=0.9\textwidth]{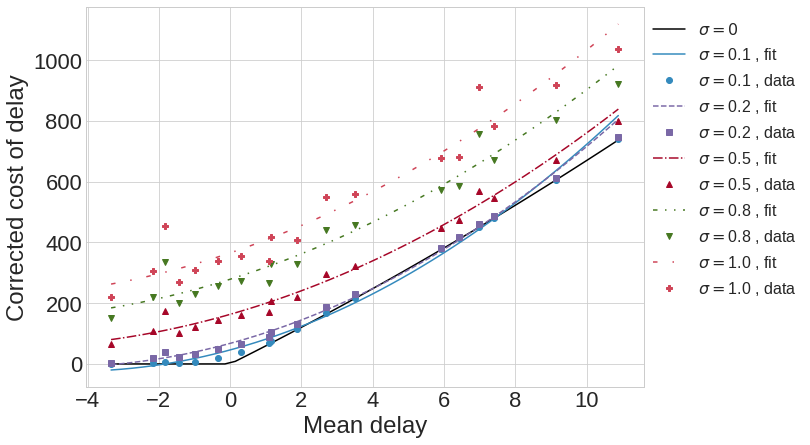}
\caption{Correction of the cost of delay function with quadratic fits.}
\label{fig:annex_cost_delay}
\end{center}
\end{figure}

Since the uncorrected cost of delay is a quadratic function, it is logical to start with such a function. As shown in Figure \ref{fig:annex_cost_delay}, the fit is quite good for such a function. However, there are several crucial issues, the first one being the overestimation of the correction for small deviations and high mean values. As a consequence, for high mean delays, the cost for low variability eventually gets larger than the costs associated with the higher value of the variabilities. As a result, the cost of delay is not a monotonically decreasing function of the variability of the departure time, which is a technical issue for the model itself, and also for its interpretation. The second point is that the corrected cost is not assured to be bigger than the uncorrected cost, as shown again by the blue line. This does not make any sense in  operational terms and thus should not occur. The third point is that quadratic functions are problematic in the far negative region of delays (not shown here), the cost will increase again at some point when the delay is sufficiently negative. This is obviously problematic when trying to find a solution for the implicit equation of delay, since the demand part is not monotonically decreasing anymore (see \ref{annex:implicit} for more details). Finally, the simple quadratic function is not required to be positive for all values of the mean delays. This is an issue since it is assumed, in accordance with the literature and our experience, that negative delays (early arrivals) do not typically represent a gain for the airlines. As a consequence, a better function is sought which has all the required properties. Using the uncorrected function as a baseline for the new function, we used:
$$
f(x) = \frac{1}{2}\left(1 - \tanh\left(\frac{x}{s}\right)\right)  \left(c + d e^{fx}\right) + \frac{1}{2}\left(1 + \tanh\left(\frac{x}{s}\right)\right) c_d (x),
$$
where $c_d$ is the initial, uncorrected cost of delay function. This function allows us to pass smoothly from the initial cost function at high mean delay down to a new exponential function at low mean delay. The transition is made smoothly thanks to a hyperbolic tangent. The results are shown in Figure \ref{fig:correction}. 

\section{Implicit equation of delay}
\label{annex:implicit}

One important feature of the model is that it is self-consistent for the delays, i.e. the delays in the output are exactly the right level to match the actual traffic, which in turn sets the average delay through the delay-capacity relationship. In other words, setting a distribution of delays fixes the actual traffic through the use of probability $P_A$, which in turn fixes the delay at each hour of the day through the capacity-delay relationship. In order to solve this circular issue, an implicit equation needs to be solved. From Equation \ref{eq:P_A} we have:
\begin{equation}
P_A(\delay) = \frac{2}{1+e^{c_d(\delay)/s}}
\label{eq:implicit}
\end{equation}
and inverting Equation \ref{eq:delay}, knowing that $T = P_A \beta$, yields:
\begin{equation}
P_A(\delay) = \frac{C}{\beta}\ln \left(\frac{\delay}{120} + cc \right).
\label{eq:implicit2}
\end{equation}
The implicit equation is solved when both expressions are equal. This equation does not have an analytical solution, but is trivial to solve numerically. Indeed, the term in Equation \ref{eq:implicit} is monotonically decreasing, whereas the term in Equation \ref{eq:implicit2} is strictly monotonically increasing and spans $(-\infty, +\infty )$. Moreover, both functions are continuous. As a consequence, there is always a unique solution to the implicit equation, and a simple, local and scalar minimiser can find it in a very small amount of time, for example using the Brent method.

\begin{figure}[htbp]
\begin{center}
\includegraphics[width=0.75\textwidth]{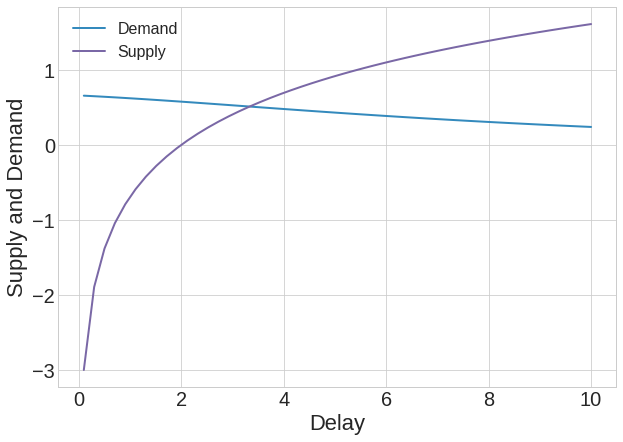}
\caption{Illustration resolution of the implicit equation of delay.}
\label{fig:implicit}
\end{center}
\end{figure}

It is interesting to realise that these two equations can be reinterpreted in terms of demand and supply curves. Indeed, equation \ref{eq:implicit} is the equivalent of a demand function, with $\delay$ playing the role of the price, and equation \ref{eq:implicit2} is the equivalent of a supply function. The `goods' exchanged can be thought as the number of flights departing from the airport. Figure \ref{fig:implicit} shows the two curves. Their intersection represents the \correctionbis{actual} delay and traffic, which is \correctionbis{equivalent} to the price and quantity of commodities actually exchanged when dealing with standard demand and supply curves.

When the problem is framed like this, some features of the model can be easily understood. For instance, the increase of the cost of delay in the demand equation drives the corresponding curve down. Conversely, when the cost of delay decreases for instance because the predictability of the departure times is higher, the demand curve is driven up. A direct consequence is that the new equilibrium point is shifted right and up on the graphic. As a consequence, the number of flights departing increases (equilibrium ordinate is higher) and the average delay increases also (equilibrium abscissa is more to the right). This shows that there exists a trade-off between the predictability and the punctuality (average delay), as explained in Section \ref{predictability}.

\section{Comparison between airports}
\label{annex:comparison}

\begin{figure}[htbp]
\begin{center}
\includegraphics[width=0.75\textwidth]{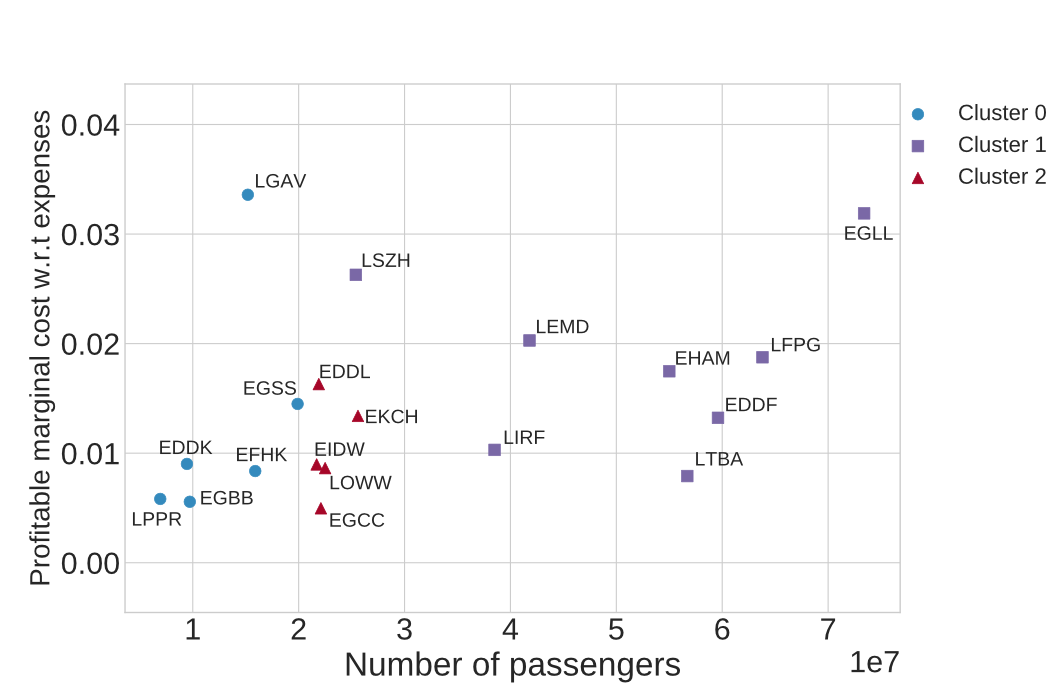}
\caption{Profitable level of marginal cost of capacity as a function of the number of passengers at the airport.}
\label{fig:alpha_pax_ratio}
\end{center}
\end{figure}

The comparison between the airports shown in Figure \ref{fig:alpha} shows that different airports have different levels of profitable marginal costs of capacity. It is also interesting to study whether with respect to their size, airports have different profitable levels of $\alpha$. Figure \ref{fig:alpha_pax_ratio} shows the ratio of the profitable level of $\alpha$ divided by the total volume of costs against the total number of passengers at the airport. Now the picture is quite different from Figure \ref{fig:alpha}, because this normalised profitable level is independent of the size of the airport. This \correctionbis{is} an important finding, because it means that larger airports are not advantaged or disadvantaged with respect to their size, but they can sustain higher capacity levels purely because they already have larger infrastructure and high costs. 

It should also be noted that whereas the size of the airport does not matter in this sense, the airports still have quite different normalised levels of profitability, from around 0.5\% up to 3\%. This could denote different management and cost efficiency levels.

\section{Sensitivity analysis}
\label{annex:sensitivity}

\begin{figure}[htbp]
\begin{center}
\includegraphics[width=0.49\textwidth]{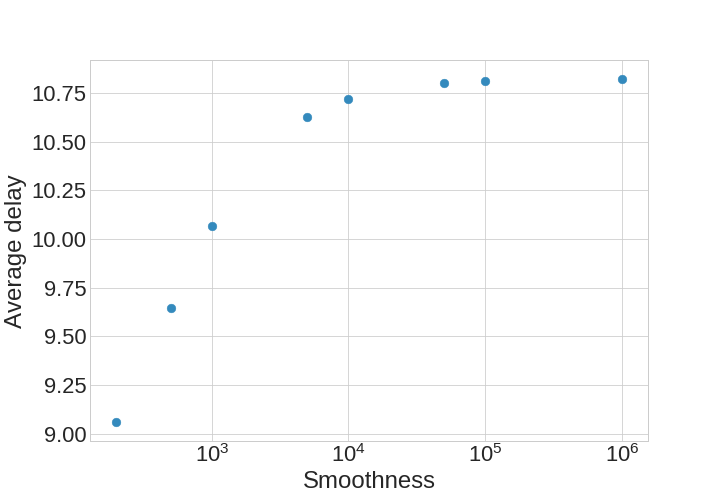}
\includegraphics[width=0.49\textwidth]{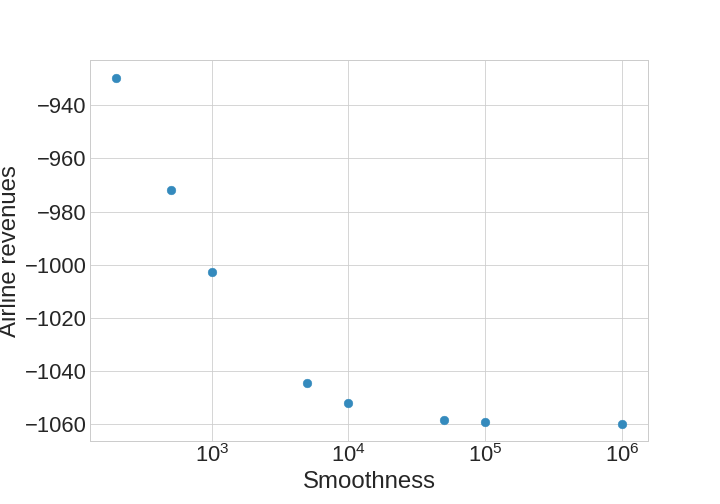}
\caption{Evolution of the average delay (left) and revenues of airlines (right) in the calibrated model for various values of the smoothness parameter $s$.}
\label{fig:sensitivity}
\end{center}
\end{figure}

In this appendix, the results of a sensitivity analysis performed on a calibrated example of a large, European hub airport are concisely shown. Since there is only one free parameter left in the model, it is simply swept to see how the calibrated parameters change. In Figure \ref{fig:sensitivity}, the evolution of the average delay in the output and the revenues of the airlines (in fact, only the cost of delay,  counted negatively) are shown. These two outputs are those of interest, all others being fixed (e.g. the revenues per passenger) or trivially related to them.
Both quantities change with the smoothness, but not remarkably. For example, delay changes from around 9 minutes per flight up to 11.7 minutes, which is a fairly narrow window, although not negligible. It is worth noting that the actual value of the delay for the calibrated airport is 9.6 minutes in the data, which means in fact that this last parameter could be calibrated to fit the average delay. This was not done for technical reasons, but in the main analysis $s=500$ was chosen, which gives a delay close to 9.5 minutes. It can thus be concluded that the results presented in the main text are sufficiently reliable with regard to the parameters.

\end{document}